Representation of solutions of linear PDE's with constant coefficients
as superpositions of solutions in lower dimensions

John R. Ockendon
OCIAM
Mathematical Institute
24-29 St. Giles'
Oxford, OX1 3LB, UK

and

Yair Zarmi
Jacob Blaustein Institute for Desert Research
Ben-Gurion University of the Negev
Sede-Boqer Campus, 84990, Israel

ABSTRACT

A unified approach to the representation of solutions of linear PDE's with constant coefficients in high dimensions in terms of solutions of the same PDE's in lower dimensions is presented. It is based on the observation that if a function of an ($N+1$)-dimensional variable ($N \geq 2$) has a convergent Fourier integral representation, then it may be written as a superposition of rotated plane waves in any number of dimensions lower than $N+1$. If that function is a solution of a linear PDE with constant coefficients, and, in addition, the equation is invariant under a spatial group of transformations (e.g., rotations, or Lorentz transformations), then the superposition is in terms of solutions of lower- (not only two-) dimensional versions of the same PDE. As examples, representations of solutions of the Laplace equation in ($N+1$) dimensions and of the wave equation in ($1+N$) dimensions ($N \geq 2$) as superpositions of solutions in lower dimensions are discussed, and simple consequences are presented.

Keywords: PDE's; Representation of solutions

MSC classification: 35C15, 35E20

PACS classification: 02.30.Jr, 41.20.Cv, 41.20.Jb

# 1. Introduction

The solutions of linear partial differential equations with constant coefficients can always be written as Fourier sums or integrals. However, such representations may not be as convenient as, say, superpositions of singular solutions. The relationship between these kinds of representations is the subject of this paper.

The task of finding solutions is often easier the lower is the dimension of the space of independent variables and, therefore, the ability to express solutions of a high-dimensional PDE in terms of its lower-dimensional solutions can be of great value. In the literature, there are several derivations, using different methods, of the representation of high-dimensional solutions of linear PDE's with constant coefficients in terms of their solutions in *two* dimensions. The earliest examples of relations between solutions in a high dimension to lower ones are in Whittaker's work [1], where he showed that the solutions of the Laplace equation in three dimensions and of the wave equation in (1+3) dimensions may be represented as superpositions of two-dimensional solutions of the same equations, and the intimately connected *reverse* procedure, the "method of descent" [2, 3], by which a (1+3)-dimensional solution of the wave equation generates a lower-dimensional one. The Radon transform analysis [4, 5] has also led [6] to a representation of the solution of a PDE as a superposition of solutions of the two-dimensional version of the same PDE, provided the original equation is invariant under the rotation group. A recent analysis [8], using twistor theory [9], has yielded a representation of the solution of the Laplace equation in three dimensions as an integral over a complex parameter of a function of the coordinates and of that parameter. This integral can be also interpreted as a superposition of two-dimensional solutions of the same equation (see Section 5).

Except for [8], which employs symmetry arguments, all the other cited works implicitly assume the existence of a Fourier integral representation of the solution. Whittaker's work can be interpreted in terms of the existence of such an integral; in the method of descent and in the Radon transform approach, the existence of an invertible Fourier integral representation is again assumed. In the remainder of this Section we review the results of [1]. In Section, 2 a unified approach, based solely on Fourier integrals, to the representation of solutions in high dimensions in terms of low- (not only two-) dimensional ones is presented. For a linear PDE with constant coefficients in ($N$+1) dimensions it will be shown that:

*(i)* Such representations are applicable to a wide class of PDE's in ($N$+1) dimensions ($N \geq 2$);
*(ii)* The lower-dimensional solutions need not be confined to two-dimensional ones;
*(iii)* Representations may be possible in terms of solutions in a range of lower dimensions.

The basic idea is that if the solution has a convergent Fourier integral representation, then the latter may be viewed as a superposition of rotated plane waves in any dimension, $2 \leq m \leq N+1$. This follows from the rotational invariance of the scalar product, **k**·**x**, which appears in the exponential factor in the Fourier integral. If the PDE is invariant under a $p$–dimensional rotation group, with $2 \leq p \leq N + 1$, then the superposition may be written in terms of rotated waves that are solutions of the *m*-dimensional version of the original PDE, for any $(N + 2 - p) \leq m \leq N + 1$.

In Sections 3 & 4, we present examples of representations of solutions of the wave equation in (1+$N$) dimensions and of the Laplace equation in ($N$+1) dimensions ($N \geq 2$) (including cases of spatial anisotropy) as superpositions of lower-dimensional solutions. In Section 5 we return to the previous works [6, 8] and show how points (*ii*) and (*iii*) above can be seen in these approaches.

In problem solving, it is important to use the most appropriate representation of the solution, and this will often depend on the nature of the initial and boundary conditions. In this paper, we will usually ignore boundary conditions and assume the solution exists in all space, or in a half-space, if there is a time-like variable.

## 1.1 <u>The wave equation: Superposition of rotated (1+1)-dimensional solutions</u>

In [1], it is shown that if the solution of the wave equation in (1+3) dimensions

$$u_{tt}^{(1+3)} = u_{xx}^{(1+3)} + u_{yy}^{(1+3)} + u_{zz}^{(1+3)} \tag{1.1}$$

has a uniformly convergent expansion in powers of $t,x,y,z$ in some domain around the origin

$$|t|, \sqrt{x^2 + y^2 + z^2} \le a \quad, \tag{1.2}$$

then the general solution can be expressed in the form

$$u^{(1+3)}(t,x,y,z) = \int_0^\pi \sin\theta\, d\theta \int_0^{2\pi} F[t - z';\theta,\varphi]d\varphi \quad, \tag{1.3}$$

where

$$z' = z\cos\theta - (x\cos\varphi + y\sin\varphi)\sin\theta \quad. \tag{1.4}$$

The geometrical significance of Eq. (1.3) is that the general solution in (1+3) dimensions is represented by a superposition of solutions of the (1+1)-dimensional wave equation at the point ($t$, $z'$), where $z'$ is the $z$ coordinate after a three-dimensional rotation.

The proof in [1] exploits the fact that monomials, $(t^q x^m y^n z^p)$, appearing in the power-series expansion of $u^{(1+3)}$, may be written as linear combinations of integrals of polynomials in $(t - z')$ over the angles $\theta$ and $\varphi$, i.e., coefficients in the Fourier expansion of such polynomials. These polynomials are solutions of the wave equation in (1+1) dimensions.

Although not stated in [1], the invariance of the wave equation under the Lorentz group, may be exploited in a similar analysis, leading to the conclusion that the general solution of that equation in (1+3) dimensions may be also written as

$$u^{(1+3)}(t,x,y,z) = \int_{n_0^2 - |\mathbf{n}|^2 = 1} u^{(1+1)}(t' - z';n_0,\mathbf{n})d\Omega_{1+3} \quad, \tag{1.5}$$

where $\Omega_{1+3}$ denotes the parameters of the Lorentz transformation that transforms the point ($t'$, $z'$) to ($t$, $x$, $y$,$z$), and ($n_0$,$\mathbf{n}$) is a four-dimensional unit vector under the metric (1,-1,-1,-1). Application to other dimensions is obvious. For instance, the (1+2)-dimensional version of Eq. (1.3) is

$$u^{(1+2)}(t,x,y) = \int_0^{2\pi} u^{(1+1)}(t - x';\varphi)d\varphi \quad. \tag{1.6}$$

In the "method of descent" [2, 3], a solution of the wave equation in (1+3) dimensions is related to a (1+2)-dimensional one, and then to the D'Alambert solution in (1+1) dimensions. For instance, consider the retarded potential representation of the (1+3)-dimensional solution

$$u^{(1+3)}(t,\mathbf{x}) = \int_{-\infty}^{\infty} \frac{\delta(t - |\mathbf{x} - \mathbf{x}'|)}{|\mathbf{x} - \mathbf{x}'|} \psi(\mathbf{x}') d^3\mathbf{x}' \quad , \tag{1.7}$$

where $\mathbf{x}, \mathbf{x}' \in R^3$ and $u(t = 0, \mathbf{x}) = 0$, $u_t^{(1+3)}(t = 0, \mathbf{x}) = 4\pi \psi(\mathbf{x})$. Eq. (1.7) may be cast in the form

$$u^{(1+3)}(t,\mathbf{x}) = t \int_0^{2\pi} d\varphi \int_0^{\pi} \psi(x + t\sin\theta\cos\varphi, y + t\sin\theta\sin\varphi, z + t\cos\varphi) \sin\theta \, d\theta \quad . \tag{1.8}$$

When $\psi$ in Eq. (1.7) is independent of $z$, the solution for a two-dimensional initial value problem with $u_t^{(1+2)}(t = 0, \mathbf{x}) = 2\pi \psi(\mathbf{x})$, $\mathbf{x} = (x,y)$, is obtained as

$$u^{(1+2)}(t,\mathbf{x}) = \int_{|\mathbf{x} - \mathbf{r}| \le t} \frac{\psi(\mathbf{x}')}{\sqrt{t^2 - |\mathbf{x} - \mathbf{x}'|^2}} d^2\mathbf{x}' \quad , \tag{1.9}$$

which is singular on the light cone, $|\mathbf{x} - \mathbf{x}'| = t$, whereas such a singularity is not apparent in Eq. (1.6). Eqs. (1.7) and (1.9) illustrate distinctions between even and odd spatial dimensions, which are not immediately apparent in Eqs. (1.3) and (1,6).

## 1.2  The Laplace equation: Superposition of rotated two-dimensional solutions

If the solution, $V_3(x,y,z)$, of the Laplace equation in three dimensions

$$\Delta V_3(x,y,z) = 0 \tag{1.10}$$

has a uniformly convergent expansion in powers of $x$, $y$, $z$ in

$$r = \sqrt{x^2 + y^2 + z^2} \le a \quad , \tag{1.11}$$

then it may be written in the form [1]

$$V_3(x,y,z) = \int_0^{2\pi} F(z - ix\cos\varphi - iy\sin\varphi; \varphi) d\varphi \quad . \tag{1.12}$$

The geometrical interpretation of Eq. (1.12) is pointed out in [1]. The quantity $x' = x\cos\varphi + y\sin\varphi$ is the result of a rotation by an angle $\varphi$ in the $x - y$ plane:

$$\begin{pmatrix} x \\ y \end{pmatrix} \to \begin{pmatrix} x' \\ y' \end{pmatrix} = \begin{pmatrix} x\cos\varphi + y\sin\varphi \\ -x\sin\varphi + y\cos\varphi \end{pmatrix} \quad . \tag{1.13}$$

Thus, the solution of the three-dimensional Laplace equation may be viewed as a superposition of solutions (possibly dependent on $\varphi$ as a parameter) of the Laplace equation in two dimensions, which are then rotated back to the three-dimensional point $(x, y, z)$:

$$V_2(x',z) = F(z - ix';\varphi) \ . \tag{1.14}$$

The proof in [1] exploits the fact that monomials, $(x^m y^n z^p)$, appearing in the expansion of $V_3$, may be expressed in terms of integrals over the angle $\varphi$ of polynomials in the quantity $(z - ix')$. These polynomials are solutions of Eq. (1.10). Although not stated in [1], the same monomials may be also expressed in terms of integrals over the angles $\theta$ and $\varphi$ of polynomials in

$$z' - ix' = (z - i(x\cos\varphi + y\sin\varphi))\exp(-i\theta) \ . \tag{1.15}$$

In Eq. (1.15) $z'$ and $x'$ are the results of a *three*-dimensional rotation, given by

$$z' = z\cos\theta - (x\cos\varphi + y\sin\varphi)\sin\theta \qquad x' = z\sin\theta + (x\cos\varphi + y\sin\varphi)\cos\theta. \tag{1.16}$$

Consequently, the invariance of the Laplacian under three-dimensional rotations leads to the conclusion that the integral over rotations about a fixed axis of Eq. (1.12) may be replaced by a superposition of values of $V_2$, computed at all points $(x', z')$ in the $x$–$z$ plane, connected to the point $(x, y, z)$ by a *three*-dimensional rotation, $R(\Omega_3)$, with $\Omega_3 = (\theta, \varphi)$:

$$V_3(\mathbf{r}) = \int_0^\pi \sin\theta \, d\theta \int_0^{2\pi} \int V_2(z' - ix';\theta,\varphi) d\varphi = \int_{|\mathbf{n}|=1} V_2(z' - ix';\mathbf{n}) d\Omega_3 \ . \tag{1.17}$$

We conclude this Section with the following comments.

*(i)* The analysis in [1] suggests the generalisation of Eqs. (1.3) and (1.12) to $N+1$ dimensions, $N > 3$: The solution may be represented as a superposition of rotated two-dimensional solutions in terms of an integral over the $(N–2)$ angles of the $N$-dimensional rotation group. This is obtained by showing that monomials, $\left(x_1^{n_1} x_2^{n_2} \cdots x_N^{n_N} x_{N+1}^{n_{N+1}}\right)$, which appear in the power expansion of the solution, may be represented as integrals over the $(N–2)$ angles of polynomials in $x_{N+1} - x_1'$, with $x_1'$ obtained by a rotation in the $N$-dimensional subspace.

*(ii)* If, instead, one considers monomials in $\left(\left(x_1^{n_1} x_2^{n_2} \cdots x_{N-m}^{n_{N-m}}\right) x_{N+1}^{n_{N+1}}\right)$, $m < N$, and expresses them as integrals over polynomials in $x_{N+1} - x_1'$, with $x_1'$ now obtained by a rotation in the $(N–m)$-dimensional subspace, then a representation in terms of solutions in $m+2$ dimensions ($x_1'$, $x_{N+1}$ and the remaining $m$ coordinates) is obtained.

*(iii)* Experience with other (e.g., Fourier) expansions may lead one to guess that results like Eq. (1.12) ought to be of the form

$$V_3(z,\mathbf{r}) = \int_0^{2\pi} F(z - ix\cos\varphi - iy\sin\varphi) w(\varphi) d\varphi \ , \tag{1.18}$$

where $w(\varphi)$ is the most general weight function. However, the various approaches reviewed in this paper show that such superpositions do not lead to general solutions.

(*iv*) The analogue of Eq. (1.9) is the familiar "source" representation [2. 3],

$$V_3(x,y,z) = \int_{-\infty}^{\infty}\int_{-\infty}^{\infty} \frac{g(x',y')}{\sqrt{z^2 + (x-x')^2 + (y-y')^2}} dx' dy' \quad , \qquad (1.19)$$

which is singular on $r = 0$ whenever $g \neq 0$.

## 2. A unified approach: Fourier integral analysis of PDE's

In this Section, the representation of solutions of a high-dimensional linear PDE with constant coefficients as superpositions of lower-dimensional solutions of the same PDE is shown to be a simple consequence of the assumption that the solution has a convergent Fourier integral representation (not required to be invertible). The dimensionality of the solutions used in the superposition may assume a range of values, 2 being the lowest. The results of previous works, discussed in Section 1 and, later on, in Section 5, emerge naturally.

### 2.1 Fourier integrals as superpositions of lower-dimensional projections

The Fourier integral representation of a function, $f(\mathbf{x})$, $\mathbf{x} \in R^N$ is written as[*]

$$f(\mathbf{x}) = \frac{1}{(2\pi)^{N/2}} \int_{-\infty}^{\infty} \tilde{f}(\mathbf{k}) e^{i\mathbf{k}\cdot\mathbf{x}} d^N\mathbf{k} \quad . \qquad (2.1)$$

The r.h.s. of Eq. (2.1) is usually interpreted as a superposition of plane waves in $N$ dimensions. The rotational invariance of the phase, $\mathbf{k}\cdot\mathbf{x}$, in the Fourier integral leads to an alternative interpretation. Let $R(\mathbf{k})$ be the rotation that rotates the vector $\mathbf{k}$ into the direction of, say, the first coordinate:

$$\begin{aligned}\mathbf{k} = R(\mathbf{k})\tilde{\mathbf{k}} \quad , & \qquad \tilde{\mathbf{k}} = (k,0,...,0) \\ \Rightarrow \quad \mathbf{k}\cdot\mathbf{x} = k x_1' \quad , & \quad \mathbf{x}' = R^{-1}\mathbf{x}\end{aligned} \qquad (2.2)$$

Eq. (2.1) becomes a superposition of *one-dimensional* waves propagating along the $x_1'$-axis, rotated by the $N$-dimensional rotation, which transforms the vector $\tilde{\mathbf{k}}$ into $\mathbf{k}$ (or $\mathbf{x}$ into $\mathbf{x}'$):

---

[*] Although we will not need Fourier inversion in our approach, we adopt the convention that

$$\tilde{f}(\mathbf{k}) = \frac{1}{(2\pi)^{N/2}} \int_{-\infty}^{\infty} f(\mathbf{x}) e^{-i\mathbf{k}\cdot\mathbf{x}} d^N\mathbf{x}$$

$$f(\mathbf{x}) = \frac{1}{(2\pi)^{N/2}} \int_{-\infty}^{\infty} \tilde{f}(\mathbf{k}) e^{ikx_1'} \, d^N\mathbf{k} \quad . \tag{2.3}$$

Eq. (2.3) may be generalised to a superposition in terms of waves in *any* dimension, $m+1$, with $0 \leq m \leq N$. To see this, we split the $N$-dimensional integration in Eq. (2.1) as follows:

$$f(\mathbf{x}) = \frac{1}{(2\pi)^{N/2}} \int_{-\infty}^{\infty} e^{i(\hat{\mathbf{k}} \cdot \hat{\mathbf{x}} + \bar{\mathbf{k}} \cdot \bar{\mathbf{x}})} \tilde{f}(\hat{\mathbf{k}}, \bar{\mathbf{k}}) d^{N-m}\hat{\mathbf{k}} \, d^m\bar{\mathbf{k}} \quad , \tag{2.4}$$

where

$$\mathbf{x} = (\hat{\mathbf{x}}, \bar{\mathbf{x}}), \quad \mathbf{k} = (\hat{\mathbf{k}}, \bar{\mathbf{k}}) \quad ,$$
$$\hat{\mathbf{x}} = (x_1, \ldots x_{N-m}), \quad \bar{\mathbf{x}} = (x_{N-m+1}, \ldots x_N) \quad \hat{\mathbf{k}} = (k_1, \ldots, k_{N-m}), \quad \bar{\mathbf{k}} = (k_{N-m+1}, \ldots, k_N) \quad . \tag{2.5}$$

Employing a rotation like Eq. (2.2) in $\hat{\mathbf{q}}, \hat{\mathbf{x}} \in R^{N-m}$, one obtains the desired superposition:

$$f(\mathbf{x}) = \frac{1}{(2\pi)^{N/2}} \int_{-\infty}^{\infty} e^{i(\hat{k} x_1' + \bar{\mathbf{k}} \cdot \bar{\mathbf{x}})} \tilde{f}(\hat{\mathbf{k}}, \bar{\mathbf{k}}) d^{N-m}\hat{\mathbf{k}} \, d^m\bar{\mathbf{k}} \quad . \tag{2.6}$$

## 2.2 **Fourier integral analysis of solutions of PDE's**

Consider a linear PDE with constant coefficients in $(N+1)$ dimensions:

$$P(\partial/\partial x_i) u(x_{N+1}, \mathbf{x}) = 0 \quad , \qquad \mathbf{x} \in R^N \quad , \qquad 1 \leq i \leq N+1 \quad . \tag{2.7}$$

As usual, we write the solution as a formal Fourier integral

$$u(x_{N+1}, \mathbf{x}) = \int_{-\infty}^{\infty} e^{i(k_{N+1} x_{N+1} + \mathbf{k} \cdot \mathbf{x})} \tilde{u}(k_{N+1}, \mathbf{k}) d^N\mathbf{k} \, dk_{N+1} \quad . \tag{2.8}$$

In Eqs. (2.7) and (2.8), one coordinate has been written out explicitly for reasons that will become apparent in the subsequent analysis.

The vector $\mathbf{q} = (k_{N+1}, \mathbf{k})$ satisfies the constraint

$$P(i\mathbf{q}) = 0 \quad , \tag{2.9}$$

which, if $x_{N+1}$ is the time variable, is the dispersion relation. In general, Eq. (2.9) determines one component of $\mathbf{q}$, say, $k_{N+1}$, which may be complex, in terms of the $N$ remaining components. For convenience, we take these components to be real, although, in practice, they may be complex as a result of contour deformation. Assuming that the integral spanned by $\mathbf{k} \in R^N$,

$$u(x_{N+1}, \mathbf{x}) = \frac{1}{(2\pi)^{N/2}} \int_{-\infty}^{\infty} e^{i(k_{N+1}(\mathbf{k}) x_{N+1} + \mathbf{k} \cdot \mathbf{x})} f(k_{N+1}(\mathbf{k}), \mathbf{k}) d^N\mathbf{k} \quad , \tag{2.10}$$

where $x_{N+1} \in R$, $\mathbf{x} \in R^N$, exists, then this is a solution of Eq. (2.7). In principle, physical requirements, and, perhaps, data on $x_{N+1} = 0$ select a branch of the solution of Eq. (2.9) for $k_{N+1}(\mathbf{k})$ and determine the function $f(k_{N+1}, \mathbf{k})$.

The standard interpretation of Eq. (2.10) is that of a superposition of plane waves in the $N$-dimensional subspace spanned by $\mathbf{x}$, their amplitude varying as a function of $x_{N+1}$. Resorting to the rotation, Eq. (2.2), and exploiting the rotational invariance of the phase factor, $\mathbf{k} \cdot \mathbf{x}$, in the Fourier integral, an interpretation in terms of waves in *two dimensions* emerges as

$$u(x_{N+1}, \mathbf{x}) = \frac{1}{(2\pi)^{N/2}} \int_{-\infty}^{\infty} e^{i(k_{N+1}(\mathbf{k}) x_{N+1} + k x_1')} f(k_{N+1}(\mathbf{k}), \mathbf{k}) d^N \mathbf{k} \quad . \tag{2.11}$$

If the operator $P(\partial/\partial x_i)$ of Eq. (2.7) is invariant under the rotation group in $\mathbf{x} \in R^N$, then $k_{n+1}(\mathbf{k})$, the solution of Eq. (2.9), is a function of $k = |\mathbf{k}|$ only. Then the plane waves in Eq. (2.12) are solutions of the two-dimensional version of Eq. (2.7). This is illustrated by the equation for the electrostatic potential in an anisotropic medium:

$$\left\{ \frac{\partial^2}{\partial z^2} + \alpha \frac{\partial^2}{\partial x^2} + \beta \frac{\partial^2}{\partial y^2} \right\} u(x, y, z) = 0 \qquad \alpha > 0, \beta > 0, z > 0 \quad , \tag{2.12}$$

where we have chosen $x_{N+1} = z$. Eq. (2.9) now becomes:

$$P(i\mathbf{q}) = -k_z(\mathbf{k})^2 - \alpha k_x^2 - \beta k_y^2 = 0 \quad . \tag{2.13}$$

To obtain a convergent integral representation for $z > 0$, we choose the root

$$k_z(\mathbf{k}) = +i\sqrt{\alpha k_x^2 + \beta k_y^2} \quad . \tag{2.14}$$

When $\beta = \alpha$, one has $q_z(\mathbf{k}) = \alpha k$, which is invariant under rotations in the $x$–$y$ plane. Then $e^{i(k_z(\mathbf{k}) z + k x')}$ [where $x'$ is the rotated variable, see Eq. (1.13)] is a solution of the two-dimensional version of Eq. (2.12). Hence, using Eq. (2.11), the solution of Eq. (2.12) may be written as

$$u(x, y, z) = \frac{1}{(2\pi)} \int_{-\infty}^{\infty} e^{k(-\alpha z + i x_1')} \tilde{u}(k_x, k_y) d^2 \mathbf{k} =$$

$$\int_0^{2\pi} \left\{ \frac{1}{(2\pi)} \int_0^{\infty} e^{k(-\alpha z + i x_1')} \tilde{u}(k \cos\theta, k \sin\theta) k\, dk \right\} d\theta = \int_0^{2\pi} F(-\alpha z + i x_1'; \theta) d\theta \quad , \tag{2.15}$$

which is a simple extension of Whittaker's result, Eq. (1.12); it has the form of a superposition of solutions of Eq. (2.12) in two dimensions, rotated by all single-axis rotations that transform the point $(x,y,z)$, at which the potential is computed, to the point $(x',0,z)$. For $\alpha \neq \beta$, $k_z(\mathbf{k})$ is not invariant under rotations in the $(k_x, k_y)$-plane and $e^{i(k_z(\mathbf{k}) z + k x')}$ is then not a solution of the two-dimensional version of Eq. (2.12).

9Returning to the general case, if $P(\partial/\partial x_i)$ is invariant under a transformation group in $(N+1)$ dimensions, then this wider group may be used instead of the *N*-dimensional one. As a result, the solution at $\mathbf{x} \in R^{N+1}$ may be written as a superposition of two-dimensional solutions of Eq. (2.7) at a point, $(x_1', x_{N+1}')$, in the $x_1$–$x_{N+1}$ plane, connected to $(x_{N+1}, \mathbf{x})$ by an $(N+1)$-dimensional transformation. This is the $(N+1)$-dimensional version of Eqs. (1.5) and (1.17). For instance, the Laplace and wave equations are invariant under $(N+1)$-dimensional rotations and $(1+N)$-dimensional Lorentz transformations, respectively. Hence, the phase in the Fourier integral for the solutions of the Laplace equation, $(-k x_{N+1} + i \mathbf{k} \cdot \mathbf{x})$, is invariant under $(N+1)$-dimensional rotations, and for the wave equation, $(\pm i k t + i \mathbf{k} \cdot \mathbf{x})$ is invariant under $(N+1)$-dimensional Lorentz transformations.

These two situations are extreme cases of problems where Eq. (2.11) can be generalised to a superposition in terms of plane waves in $(N-m+2)$ dimensions, for any $1 \leq m \leq N$. To do this, we employ the decomposition of Eq. (2.5), and apply a rotation in the subspace $\overline{\mathbf{k}} \in R^m$:

$$u(x_{N+1}, \mathbf{x}) = \frac{1}{(2\pi)^{N/2}} \int_{-\infty}^{\infty} e^{i(k_{N+1}(\mathbf{k}) x_{N+1} + \overline{k}\, \overline{x}'_1 + \hat{\mathbf{k}} \cdot \hat{\mathbf{x}})} f(k_{N+1}(\mathbf{k}), \mathbf{k}) d^N \mathbf{k} \quad . \tag{2.16}$$

If we now add the assumption that $P(\partial/\partial x_i)$ is invariant under rotations in a *p*-dimensional subspace, $\hat{\mathbf{x}} \in R^p$, $p \leq N$, it then pays to write the solution as

$$u(x_{N+1}, \mathbf{x}) = \frac{1}{(2\pi)^{N/2}} \int_{-\infty}^{\infty} e^{i(k_{N+1}(\mathbf{k}) x_{N+1} + \hat{\mathbf{k}} \cdot \hat{\mathbf{x}} + \overline{\mathbf{k}} \cdot \overline{\mathbf{x}})} \tilde{f}(\hat{\mathbf{k}}, \overline{\mathbf{k}}) d^p \hat{\mathbf{k}} \, d^{N-p} \overline{\mathbf{k}} \quad , \tag{2.17}$$

where

$$\mathbf{x} = (\hat{\mathbf{x}}, \overline{\mathbf{x}}), \quad \mathbf{k} = (\hat{\mathbf{k}}, \overline{\mathbf{k}}) \quad ,$$
$$\hat{\mathbf{x}} = (x_1, \ldots x_p), \quad \overline{\mathbf{x}} = (x_{p+1}, \ldots x_N) \quad , \hat{\mathbf{k}} = (k_1, \ldots k_p), \quad \overline{\mathbf{k}} = (k_{p+1}, \ldots k_N) \quad . \tag{2.18}$$

Now, decompose $\hat{\mathbf{x}}$ into $\hat{\mathbf{x}} = (\boldsymbol{\eta}, \boldsymbol{\sigma})$ with $\boldsymbol{\sigma} \in R^m$, $m \leq p$, and apply a rotation such as Eq. (2.2) to the subspace spanned by $\boldsymbol{\sigma}$. The solution is then represented by a superposition of wave solutions in $(N-m+2)$ dimensions, with $(N-p+2) \leq (N-m+2) \leq (N+1)$:

$$u(x_{N+1}, \mathbf{x}) = \frac{1}{(2\pi)^{N/2}} \int_{-\infty}^{\infty} e^{i(k_{N+1} x_{N+1} + k_\sigma \sigma'_1 + \mathbf{k}_\eta \cdot \boldsymbol{\eta} + \overline{\mathbf{k}} \cdot \overline{\mathbf{x}})} \tilde{f}(\hat{\mathbf{k}}, \overline{\mathbf{k}}) d^m \boldsymbol{\sigma}\, d^{p-m} \boldsymbol{\eta}\, d^{N-p} \overline{\mathbf{k}} \quad . \tag{2.19}$$

The previous results are summarized as follows:

• If $P(\partial/\partial x_i)$ is invariant under rotations in a *p*-dimensional subspace, $1 \leq p \leq N$, then the solution of Eq. (2.7) in $N+1$ dimensions may be written as a superposition of solutions of versions of Eq. (1.1) in $(N-m+2)$ dimensions, for any $1 \leq m \leq p$. The integration is then performed over all elements of the rotation group in *m* dimensions.

- If $P(\partial/\partial x_i)$ is invariant under rotations (or Lorentz transformations) in the *full* (N+1)-dimensional space, then the representation in terms of two-dimensional solutions may be an integral over either the *N*-dimensional rotation group, or the full *(N+1)*-dimensional group.

In light of this discussion, let us now return to the examples of Section 1.

## 3. Wave equation in 1 + *N* dimensions

We have seen in Section 1 that there are several representations of solutions of the wave equation in high dimensions that are based on the assumption that the solution has a Fourier integral representation, although this is not obvious from the final form of the representation. We now examine some of these representations, and show how they can be transformed into superpositions of two-dimensional solutions.

Consider the wave equation in 1 + *N* dimensions

$$u^{(1+N)}_{tt}(t,\mathbf{x}) = \Delta u^{(1+N)}(t,\mathbf{x}) \qquad \mathbf{x} \in R^N \quad , \tag{3.1}$$

with $u^{(1+N)}(t=0,\mathbf{x}) = 0$, $u^{(1+N)}_t(t=0,\mathbf{x}) = \psi(\mathbf{x})$. For odd $N \geq 3$, the general solution for $u^{(1+N)}(t,\mathbf{x})$ may be written in the form

$$u^{(1+N)}(t,\mathbf{x}) = (-1)^{(N-1)/2} \frac{\partial^{N-2}}{\partial t^{N-2}} v(t,\mathbf{x}) \quad , \tag{3.2}$$

where

$$v(t,\mathbf{x}) = \frac{(-1)^{(N-1)/2}}{(N-2)!} \int_0^t \left(t^2 - \xi^2\right)^{(N-3)/2} \xi Q(\mathbf{x},\xi) d\xi \quad , \tag{3.3}$$

and

$$Q(\mathbf{x},\xi) = \frac{1}{\Omega_N} \int_{|\mathbf{n}|=1} \psi(\mathbf{x} + \xi\mathbf{n}) d\Omega_N \quad ; \tag{3.4}$$

in Eq. (3.4), $\Omega_N$ is the surface area element of the unit sphere in *N* dimensions and **n** is a unit vector normal to the surface of that sphere. Eqs. (3.2)–(3.4) may be derived either by generalising the retarded potential representation to higher dimensions and using the method of descent [2], or by assuming that the solution has a Fourier integral representation (see [3], where Eq. (3.2) is generalised to even values of *N*). Now consider the implication of this result when $u^{(1+N)}(t,\mathbf{x})$ is assumed to have a Fourier-integral representation

$$u^{(1+N)}(t,\mathbf{x}) = \int_{-\infty}^{\infty} d^N\mathbf{k}\, e^{i\mathbf{k}\cdot\mathbf{x}} A(\mathbf{k}) \sin kt \quad . \tag{3.5}$$

Setting aside questions of convergence, this means that $v(t,\mathbf{x})$ can be formally represented as

$$v(t,\mathbf{x}) = \int_{-\infty}^{\infty} \frac{1}{k^{N-2}} d^N\mathbf{k}\, A(\mathbf{k})\, e^{i\mathbf{k}\cdot\mathbf{x}} \cos kt =$$

$$\int_{|\mathbf{n}|=1} d\Omega_N \left( \int_0^{\infty} k\, dk\, \tfrac{1}{2}\left[ A(k,\mathbf{n})e^{ik(t-\mathbf{n}\cdot\mathbf{x})} + A(k,-\mathbf{n})e^{-ik(t-\mathbf{n}\cdot\mathbf{x})} \right] \right) = \int_{|\mathbf{n}|=1} d\Omega_N\, G(t-\mathbf{n}\cdot\mathbf{x};\mathbf{n}) \quad . \quad (3.6)$$

As a result, the general solution, Eq. (3.2), of the wave equation in 1+$N$ dimensions may be written as a superposition of solutions of the equation in 1+1 dimensions with the $x$-axis rotated by $N$-dimensional rotations:

$$u^{(1+N)}(t,\mathbf{x}) = (-1)^{(N-1)/2} \frac{\partial^{N-2}}{\partial t^{N-2}} v(t,\mathbf{x}) = \int_{\Omega_N} d\Omega_N\, F(t-\mathbf{n}\cdot\mathbf{x};\mathbf{n}) \quad , \tag{3.7}$$

where $F(t-x';\mathbf{n})$ ($x'=\mathbf{n}\cdot\mathbf{x}$) is a solution of the wave equation in 1+1 dimensions in a rotated frame (the unit vector $\mathbf{n}$ is the first row of an $N$-dimensional rotation matrix), given by:

$$F(\rho;\mathbf{n}) = (-1)^{(N-1)/2} \frac{\partial^{N-2}}{\partial \rho^{N-2}} G(\rho;\mathbf{n}) \quad . \tag{3.8}$$

**Examples**

(*i*) Solutions in (1+3) dimensions

We can quickly see that Whittaker's result, Eq. (1.3), is readily obtained from Eq. (3.7) because

$$u^{(1+3)}(t,\mathbf{x}) = \frac{1}{(2\pi)^{3/2}} \int_{-\infty}^{\infty} e^{i(kt-\mathbf{k}\cdot\mathbf{x})} \tilde{u}^{(1+3)}(0,\mathbf{k})\, d^3\mathbf{k} = \int_0^{2\pi} d\varphi \int_0^{\pi} \sin\theta\, d\theta\, F(t-x';\mathbf{n})\Big|_{\substack{x'=\mathbf{n}\cdot\mathbf{x}\\ \mathbf{n}=\mathbf{k}/k}} \quad , \tag{3.9}$$

where, here and below, we have replaced our usual wave number $\mathbf{k}$ by $-\mathbf{k}$ for convenience, and

$$F(t-x';\mathbf{n})\Big|_{\substack{x'=\mathbf{n}\cdot\mathbf{r}\\ \mathbf{n}=\mathbf{k}/k}} = \frac{1}{(2\pi)^{3/2}} \int_0^{\infty} e^{ik(t-x')} \tilde{u}^{(1+3)}(0,\mathbf{k})\, k^2\, dk \quad . \tag{3.10}$$

Note that Eq. (3.9) can be obtained, and also the relation between the source function in the retarded potential representation, Eq. (1.7), and $F(t-x';\mathbf{n})$ of Eq. (3.9) can be elucidated as follows. Fourier-expanding $\psi(\mathbf{x}')$ in Eq. (1.7),

$$\psi(\mathbf{x}') = \frac{1}{(2\pi)^{3/2}} \int_{-\infty}^{\infty} \tilde{\psi}(\mathbf{k})\, e^{-i\mathbf{k}\cdot\mathbf{x}'} d^3\mathbf{k} \quad , \quad \psi(\mathbf{x}') \in R, \mathbf{k} \in R^3 \Rightarrow \tilde{\psi}(-\mathbf{k}) = \tilde{\psi}(\mathbf{k})^* \quad , \tag{3.11}$$

we obtain, with $\tilde{\psi}(\mathbf{k}) \equiv \tilde{\psi}(k\mathbf{n})$, $\mathbf{n} = (\sin\theta\cos\varphi, \sin\theta\sin\varphi, \cos\theta)$ and $\boldsymbol{\rho} = \mathbf{x}' - \mathbf{x}$,

$$u^{(1+3)}(t,\mathbf{x}) = \frac{1}{(2\pi)^{1/2}} \int_{-\infty}^{\infty} \tilde{\psi}(\mathbf{k}) e^{-i\mathbf{k}\cdot\mathbf{x}} d^3\mathbf{k} \int_0^{\pi} \sin\theta \, d\theta \int_{-\infty}^{\infty} \delta(t-\rho) e^{-ik\rho\cos\theta} \rho \, d\rho =$$

$$\frac{1}{(2\pi)^{1/2}} \int_{-\infty}^{\infty} d^3\mathbf{k} \frac{\tilde{\psi}(k\mathbf{n})}{ik} \left\{ e^{ik(t-\mathbf{n}\cdot\mathbf{x})} - e^{-ik(t+\mathbf{n}\cdot\mathbf{x})} \right\} \quad . \tag{3.12}$$

Defining $z' = \mathbf{n}\cdot\mathbf{x} = x\sin\theta\cos\varphi + y\sin\theta\sin\varphi + z\cos\theta$, Eq. (1.3) is retrieved, with

$$F(t-z';\mathbf{n}) = \int_0^{\infty} \frac{k\,dk}{(2\pi)^{1/2} i} \left\{ \tilde{\psi}(k\mathbf{n}) e^{ik(t-z')} - \tilde{\psi}(k\mathbf{n})^* e^{-ik(t-z')} \right\} \quad . \tag{3.13}$$

We end the discussion of the connection between the retarded potential and plane-wave representations in the (1+3)-dimensional case by providing some explicit details, which are similar to results to be presented in (*iii*) below for solutions in (1+2) dimensions. The Fourier coefficients

$$F(\xi;\mathbf{n}) = \frac{1}{\sqrt{2\pi}} \int_{-\infty}^{\infty} \tilde{F}(q;\mathbf{n}) e^{-iq\xi} dq \quad , \tag{3.14}$$

and of the source function, $\psi(\mathbf{x})$, are related by

$$\tilde{F}(q;\mathbf{n}) = \begin{cases} iq\tilde{\psi}(q,\mathbf{n})^* & q > 0 \\ \\ -iq\tilde{\psi}(-q,\mathbf{n}) & q < 0 \end{cases} . \tag{3.15}$$

Exploiting the initial data, $u_t(t=0,\mathbf{x}) = 4\pi\,\psi(\mathbf{x})$, the source function, $\psi(\mathbf{x})$, can be related to the rotated solution in 1+1 dimensions, computed at $t = 0$, by:

$$\psi(\mathbf{x}) = \frac{1}{4\pi} u_t(t=0,\mathbf{x}) = \frac{1}{4\pi} \int_0^{\pi} \sin\theta \, d\theta \int_0^{2\pi} d\varphi \, F_\xi(\xi;\mathbf{n})\Big|_{\xi=-\mathbf{n}\cdot\mathbf{x}} \quad . \tag{3.16}$$

(*ii*) Cylindrically symmetric solutions in (1+3) dimensions

Writing $x = r\cos\phi$, $y = r\sin\phi$ in the Fourier integral in Eq. (3.9) and requiring that the solution has cylindrical symmetry (i.e., is independent of $\phi$), one obtains the familiar result that

$$u^{(1+3)}(t,r,z) = \frac{1}{2\pi} \int_0^{2\pi} d\phi \left( \frac{1}{(2\pi)^{3/2}} \int_0^{\pi} \sin\theta \, d\theta \int_0^{2\pi} d\varphi \int_0^{\infty} e^{ik(t-(r\sin\theta\cos(\phi-\varphi)+z\cos\theta))} \tilde{u}^{(1+3)}(0,\mathbf{k}) k^2 \, dk \right)$$

$$= \frac{1}{2\pi} \int_0^{2\pi} u^{(1+2)}(t,x=r\cos\phi,z) d\phi \quad , \tag{3.17}$$

where

$$u^{(1+2)}(t,x,z) = \frac{1}{(2\pi)^{3/2}} \int_0^\pi \sin\theta\, d\theta \int_0^{2\pi} d\varphi \int_0^\infty e^{ik(t-(x\sin\theta + z\cos\theta))} \tilde{u}^{(1+3)}(0,\mathbf{k}) k^2\, dk \qquad (3.18)$$

is a solution of the wave equation in 1+2 dimensions. Hence, the axially symmetric solution in 1+3 dimensions is obtained as a superposition of rotated solutions in 1+2 dimensions (which, in turn, may be written as superpositions of rotated 1+1 solutions). This is a manifestation of what we said at the end of Section 2, namely, that the representation in terms of lower-dimensional solutions is not confined to two dimensions.

Setting plane-wave representations aside for the moment, exploitation of Eq. (1.9) yields another representation of a cylindrically symmetric solution of the wave equation in (1+3) dimensions as a superposition of solutions in (1 + 2) dimensions:

$$u^{(1+3)}(t,r,z) = \int_0^{2\pi} \left\{ \iint_{(z-z')^2 + (r\cos\varphi - x')^2 \leq t^2} \frac{\psi(x',z')}{\sqrt{t^2 - \{(z-z')^2 + (r\cos\varphi - x')^2\}}} dx'\, dz' \right\} d\varphi \;, \qquad (3.19)$$

where

$$u^{(1+3)}(t=0,r,z) = 0, \quad u_t^{(1+3)}(t=0,r,z) = 2\pi \int_0^{2\pi} \psi(r\cos\varphi, z) d\varphi \;. \qquad (3.20)$$

Eq. (3.19) generalizes trivially to the case of wave propagation in an anisotropic medium with cylindrical symmetry. Similarly, spherically symmetric solutions are obtained as superposition of rotated 1+1 solutions, by expressing $x$, $y$, $z$ in spherical coordinates, and averaging with respect to both angles.

(*iii*) <u>Solutions in 1+2 dimensions</u>

In (1+2) dimensions it is easy to see that the Fourier integral representation is intimately related to the singularity distribution representation of Eq. (1.9). Writing $\psi(\mathbf{x})$ as a Fourier integral

$$\psi(\mathbf{x}) = \frac{1}{2\pi}\int_{-\infty}^\infty d^2\mathbf{k}\, \tilde{\psi}(\mathbf{k}) e^{-i\mathbf{k}\cdot\mathbf{x}} \;, \qquad \psi(\mathbf{x})\in R, \mathbf{k}\in R^2 \Rightarrow \tilde{\psi}(-\mathbf{k}) = \tilde{\psi}(\mathbf{k})^* \;, \qquad (3.21)$$

and using the notation $\mathbf{n}\cdot\mathbf{x} = x\cos\varphi + y\sin\varphi$, Eq. (1.9) becomes

$$u^{(1+2)}(t,\mathbf{x}) = \frac{1}{2\pi} \int_{-\infty}^\infty d^2\mathbf{k}\, \tilde{\psi}(\mathbf{k}) e^{-i\mathbf{k}\cdot\mathbf{x}} \int_{|\boldsymbol{\rho}|\leq t} \frac{d^2\boldsymbol{\rho}}{\sqrt{t^2 - \rho^2}} e^{-i\mathbf{k}\cdot\boldsymbol{\rho}} = \int_{-\infty}^\infty d^2\mathbf{k}\, \tilde{\psi}(\mathbf{k}) e^{-i\mathbf{k}\cdot\mathbf{x}} \frac{1}{k}\sin kt$$

$$= \int_0^{2\pi} d\varphi \frac{1}{2i} \int_0^\infty dk \left[ \tilde{\psi}(k\mathbf{n}) e^{ik(t-\mathbf{n}\cdot\mathbf{x})} - \tilde{\psi}(k\mathbf{n})^* e^{-ik(t-\mathbf{n}\cdot\mathbf{x})} \right] \;. \qquad (3.22)$$

Here, introducing polar coordinates for the $\boldsymbol{\rho}$ integration, we have used the identities [10]

$$\frac{1}{2\pi}\int_0^{2\pi} e^{ik\rho\cos\theta}d\theta = J_0(k\rho), \qquad \int_0^t \frac{\rho\,d\rho}{\sqrt{t^2-\rho^2}}\frac{1}{k}J_0(k\rho) = \frac{1}{k}\sin kt \quad. \tag{3.23}$$

Thus, Eq. (1.9) may be cast in the form of a superposition of solutions of the wave equation in (1+1) dimensions:

$$u^{(1+2)}(t,\mathbf{x}) = \int_0^{2\pi} d\varphi\, F(t-\mathbf{n}\cdot\mathbf{x};\mathbf{n}) \quad, \tag{3.24}$$

where

$$F(t-\mathbf{n}\cdot\mathbf{x};\mathbf{n}) = \frac{1}{2i}\int_0^\infty dk\left[\tilde{\psi}(k\mathbf{n})e^{ik(t-\mathbf{n}\cdot\mathbf{x})} - \tilde{\psi}(k\mathbf{n})^* e^{-ik(t-\mathbf{n}\cdot\mathbf{x})}\right] \quad. \tag{3.25}$$

As in the case of solutions in (1+3) dimensions (see Eq. (3.15)), a simple relation exists between the Fourier coefficients of $F(\xi;\mathbf{n})$:

$$F(\xi;\mathbf{n}) = \frac{1}{\sqrt{2\pi}}\int_{-\infty}^\infty \tilde{F}(q;\mathbf{n})e^{-iq\xi}\,dq \quad, \tag{3.26}$$

and the Fourier coefficients of the source function, $\psi(\mathbf{x})$:

$$\tilde{F}(q;\mathbf{n}) = \begin{cases} i\sqrt{\dfrac{\pi}{2}}\, x\tilde{\psi}(q,\mathbf{n})^* & q > 0 \\ \\ -i\sqrt{\dfrac{\pi}{2}}\,\tilde{\psi}(-q,\mathbf{n}) & q < 0 \end{cases} \quad. \tag{3.27}$$

Also (see Eq. (3.16)) the source function, $\psi(\mathbf{x})$, can be related to the rotated solution in 1+1 dimensions, computed at $t = 0$, by exploiting the initial data, $u_t^{(1+2)}(t = 0,\mathbf{x}) = 2\pi\psi(\mathbf{x})$:

$$\psi(\mathbf{x}) = \frac{1}{2\pi}u_t^{(1+2)}(t=0,\mathbf{x}) = \frac{1}{2\pi}\int_0^{2\pi} d\varphi\, F_\xi(\xi;\mathbf{n})\Big|_{\xi=-\mathbf{n}\cdot\mathbf{x}} \quad. \tag{3.30}$$

## 4. Laplace equation in 3 dimensions

Reverting to the notation of Eq. (2.1), Whittaker's result, Eq. (1.12), is obtained at once if $V_3$ has a convergent Fourier expansion for $z > 0$:

$$V_3(x,y,z) = \frac{1}{2\pi}\int_{-\infty}^{\infty} d^2\mathbf{k}\, e^{-kz}\, e^{i(k_1 x + k_2 y)}\, \tilde{V}_3(0,\mathbf{k}) = \int_0^{2\pi} d\varphi \left[ \frac{1}{2\pi}\int_0^{\infty} k\, dk\, e^{-k(z - ix\cos\varphi - iy\sin\varphi)} \left(\tilde{V}_3(0,k,\varphi)\right)^* \right]$$

(4.1)

$$= \int_0^{2\pi} F(z - ix';\varphi)\, d\varphi$$

for some analytic function $F$, where $x' = x\cos\varphi + y\sin\varphi$. Clearly, $x'$ is the result of a rotation by an angle $\varphi$. Continuation to $z < 0$ is possible if $F$ is meromorphic in

$$\xi = z - ix' \quad . \tag{4.2}$$

Comment: The general solution of the 2-dimensional equation is usually written as

$$V(z,x) = h(z - ix) + g(z + ix) \quad , \tag{4.3}$$

where $h$, $g$ are analytic. For our purposes, Eq. (4.3) may be viewed [1] as a sum over all elements of the rotation group that transform the point $x$ to any point $x'$, which in one dimension are only the identity and the reflection:

$$V(z,x) = f(z - ix; \theta = 0) + f(z + ix; \theta = \pi) \quad . \tag{4.4}$$

### 4.1 Axially symmetric solution of the 3-dimensional Laplace equation

Applying the representation of a cylindrically symmetric solution in three dimensions as a superposition in terms of two-dimensional solutions yields a simple derivation of a well-known result. As in Eq. (3.17), we use cylindrical coordinates ($x = r\cos\phi$, $y = r\sin\phi$). The Fourier integral in Eq. (4.1) becomes

$$V_3(x,y,z) = \frac{1}{2\pi}\int_0^{2\pi} d\phi \left( \int_0^{2\pi} d\varphi \frac{1}{2\pi}\int_0^{\infty} e^{-k(z - ir\cos(\phi - \varphi))} \left(\tilde{V}_3(0,k,\varphi)\right)^* k^2\, dk \right)$$

$$= \frac{1}{2\pi}\int_0^{2\pi} d\phi \left( \int_0^{2\pi} d\varphi \frac{1}{2\pi}\int_0^{\infty} e^{-k(z - ir\cos\phi)} \left(\tilde{V}_3(0,k,\varphi)\right)^* k^2\, dk \right) \quad . \tag{4.5}$$

Using the identity

$$V_3(x = y = 0, z) = \int_0^{2\pi} d\varphi \frac{1}{2\pi}\int_0^{\infty} e^{-kz} \left(\tilde{V}_3(0,k,\varphi)\right)^* k^2\, dk \quad , \tag{4.6}$$

yields the well-known result:

$$V_3(r,z) = \frac{1}{\pi}\int_0^{\pi} V_3(x = y = 0, z - ir\cos\phi)\, d\phi \quad . \tag{4.7}$$

Note that the axially symmetric version of Eq. (1.19) is obtained by setting $g(x',y') = g_0(x') \cdot \delta(y')$, so that with $y = \rho \cos\varphi$, $z = \rho \sin\varphi$,

$$V_3(\rho, x) = \int_{-\infty}^{\infty} \frac{g_0(x')}{\sqrt{\rho^2 + (x - x')^2}} dx' \quad . \tag{4.8}$$

This solution is singular on $\rho = 0$ whenever $g_0 \neq 0$, whereas

$$V_3(\rho, x) = \int_{x-i\rho}^{x+i\rho} \frac{g_0(x')}{\sqrt{\rho^2 + (x - x')^2}} dx' \quad , \tag{4.9}$$

which is also an axially symmetric solution of the Laplace equation, is analytic on $\rho = 0$.

Finally, we note that the axially symmetric solution used in Eq. (3.19), namely,

$$V_3(r, t) = \int_0^{r-t} \frac{g(\tau) d\tau}{\sqrt{r^2 - (t - \tau)^2}} \quad , \tag{.4.10}$$

is singular both on $r = 0$ and $r = t$. We repeat our observation that, whereas, source-function representations may or may not have an explicit singular behaviour, there are no apparent singularities in the representation in terms of lower-dimensional solutions.

## 5. Review and extension of previous works

### 5.1 Representation of the solution as an integral over a parameter

Consider, again, the Laplace equation in 3 dimensions. Using the following transformation of variables,

$$\varphi \in [0, 2\pi] \to \omega \in (-\infty, +\infty) \quad , \quad \cos\varphi = \frac{\omega^2 - 1}{\omega^2 + 1} \quad , \quad \sin\varphi = \frac{2\omega}{\omega^2 + 1} \quad , \tag{5.1}$$

Eq. (4.1) becomes an integral over a parameter:

$$V_3(x, y, z) = \int_{-\infty}^{\infty} \frac{2}{\omega^2 + 1} F\left(z - ix\frac{\omega^2 - 1}{\omega^2 + 1} - iy\frac{2\omega}{\omega^2 + 1}; \omega\right) d\omega \equiv$$

$$\int_{-\infty}^{\infty} h\left[(\omega^2 + 1)z - i(\omega^2 - 1)x - i2\omega y; \omega\right] d\omega \quad . \tag{5.2}$$

A recent analysis [8], using twistor theory [9], has independently produced a complex plane version of Eq. (5.2). The result of [8] is that for any curve, $C$, in the complex plane,

$$V_3(x,y,z) = \int_C h\left[(\omega^2+1)z - i(\omega^2-1)x - i2\omega y; \omega\right] d\omega \tag{5.3}$$

is a solution of Eq. (1.1), the geometrical interpretation of which is again that of integration over three-dimensional rotations. To see this, write

$$\omega = u + iv, \qquad u,v \in R. \tag{5.4}$$

The argument of $f$ in Eq. (5.3) may be then written as

$$(\omega^2+1)z - i(\omega^2-1)x - i2\omega y = (|\omega|^2+1)[z'-ix'] . \tag{5.5}$$

Here $x'$ and $z'$ are the $x$ and $z$ coordinates, rotated under the following 3-dimensional rotation:

$$\begin{pmatrix} x' \\ y' \\ z' \end{pmatrix} = \frac{1}{(|\omega|^2+1)} \begin{pmatrix} u^2-v^2-1 & 2u & -2uv \\ -2u & u^2+v^2-1 & 2v \\ 2uv & 2v & u^2-v^2+1 \end{pmatrix} \begin{pmatrix} x \\ y \\ z \end{pmatrix} . \tag{5.6}$$

Integration along $C$ may be, therefore, interpreted as a superposition of two-dimensional solutions, rotated by a one-parameter subset of the 3-dimensional rotation group:

$$V_3(x,y,z) = \int_{s_1}^{s_2} h\left[(|\omega(s)|^2+1)[z'-ix']; \omega(s)\right] \frac{d\omega(s)}{ds} ds . \tag{5.7}$$

The most general solution is obtained if independent integrations are carried out over $u$ and $v$,

$$V_3(x,y,z) = \int_{-\infty}^{\infty}\int_{-\infty}^{\infty} h\left[(u^2+v^2+1)(z'-ix'); u,v\right] du\, dv , \tag{5.8}$$

or, alternatively, over the angles, $\theta$ and $\varphi$, as in Eq. (1.17).

The analogous representation in the case of the wave equation in (1+3) dimensions, involves two-parameters. In Eq. (3.9), use the following change of variables:

$$\cos\theta = \frac{\omega^2-1}{\omega^2+1}, \quad \sin\theta = \frac{2\omega}{\omega^2+1}, \quad \cos\varphi = \frac{\sigma^2-1}{\sigma^2+1}, \quad \sin\varphi = \frac{2\sigma}{\sigma^2+1} . \tag{5.9}$$

Integrating by parts with respect to $\omega$, to avoid a double pole at $\omega = \pm i$, Eq. (3.9) then leads to:

$$u(t,\mathbf{r}) = \int_{-\infty}^{\infty} d\omega \int_{-\infty}^{\infty} d\sigma\, h\left[(\omega^2+1)(\sigma^2+1)t - 2\omega(\sigma^2-1)x - 4\omega\sigma y - (\omega^2-1)(\sigma^2+1)z; \omega,\sigma\right], \tag{5.10}$$

where

$$h\left[(\omega^2+1)(\sigma^2+1)t - 2\omega(\sigma^2-1)x - 4\omega\sigma y - (\omega^2-1)(\sigma^2+1)z;\omega,\sigma\right] = \frac{2}{(\omega^2+1)(\sigma^2+1)}\frac{dF}{d\omega} \quad . \tag{5.11}$$

When the integrations over real $\omega$ and $\sigma$ in Eq. (5.10) are replaced by integrations along curves in the complex $\omega$- & $\sigma$- planes, the resulting quantities are also solutions of the 1+3 wave equation:

$$u(t,\mathbf{r}) = \int_{C_\omega} d\omega \int_{C_\sigma} d\sigma\, h\left[(\omega^2+1)(\sigma^2+1)t - 2\omega(\sigma^2-1)x - 4\omega\sigma y - (\omega^2-1)(\sigma^2+1)z;\omega,\sigma\right] \tag{5.12}$$

The argument of $h$ in Eq. (5.12) may be written as

$$(|\omega|^2+1)(|\sigma|^2+1)\{t' - x'\} \quad . \tag{5.13}$$

Here $t'$ and $x'$ are the transforms of the 4-vector $(t, x, y, z)$ into $(t', x', 0, 0)$ by the appropriate Lorentz transformation. Integration along specific curves, $C_\omega$ and $C_\sigma$, corresponds to a superposition over a two-parameter subset of Lorentz transformations.

## 5.2 **The Radon transform**

Consider the following linear PDE with constant coefficients in $(N+1)$ dimensions, assumed to be invariant under the rotation group in $\mathbf{x} \in R^N$:

$$P(\partial/\partial x_i)u(x_{N+1},\mathbf{x}) = 0 \quad , \qquad \mathbf{x} \in R^N \quad , \qquad 1 \le i \le N+1 \quad . \tag{5.14}$$

Use of the Radon transform [4, 5] has led [6] to a representation of the solution of Eq. (5.14) as a superposition of solutions of the two-dimensional version of the same PDE. Today, the Radon transform of functions of two variables [7] is an essential tool in image reconstruction, when scanning devices that measure the total intensity of, say, an electric field along parallel lines through a specimen are employed (e.g., in tomography). However, originally [4, 5], it was defined for functions of any number of real variables for other purposes, such as the smoothing of the singular behaviour of functions or distributions. The assumptions required to obtain the results of [6] are: (*i*) the solution has an invertible Fourier representation in the *N*-dimensional subspace $\mathbf{x} \in R^N$; (*ii*) the solution and a sufficient number of its derivatives vanish at infinity.

The Radon transform of $u(x_{N+1},\mathbf{x})$ in $\mathbf{x} \in R^N$ is defined as

$$\hat{u}(x_{N+1},\rho\mathbf{n}) = \int_{-\infty}^{\infty} u(x_{N+1},\mathbf{x})\delta(\rho - \mathbf{x}\cdot\mathbf{n})d^N\mathbf{x} = \int_{(\boldsymbol{\xi}\cdot\mathbf{n})=0} u(x_{N+1},\rho\mathbf{n}+\boldsymbol{\xi})d^{N-1}\boldsymbol{\xi} \quad , \tag{5.15}$$

where **n** is a unit vector in $N$ dimensions. It is easy to see that $\hat{u}(x_{N+1}, \rho\mathbf{n})$ is a solution of a PDE in two dimensions:

$$\tilde{P}(\partial_{x_{N+1}}, \partial_\rho; \mathbf{n})\hat{u}(x_{N+1}, \rho\mathbf{n}) = 0 \quad , \tag{5.16}$$

where

$$\tilde{P}(\partial_{x_{N+1}}, \partial_\rho; \mathbf{n}) = P\left(\partial_{x_{N+1}}, A(\mathbf{n})\begin{pmatrix}\partial_\rho \\ \nabla_\xi\end{pmatrix}\right)\bigg|_{\nabla_\xi \to 0} \quad , \tag{5.17}$$

and $A(\mathbf{n})$ is the $N \times N$ matrix that generates the transformation from $\mathbf{x} \in R^N$ to $(\rho, \boldsymbol{\xi})$, $\boldsymbol{\xi} \in R^{N-1}$:

$$\nabla_{\mathbf{x}} = A(\mathbf{n})\begin{pmatrix}\partial_\rho \\ \nabla_\xi\end{pmatrix} \quad . \tag{5.18}$$

If the original PDE is invariant under the rotation group in $\mathbf{x} \in R^N$, then the operator $\tilde{P}(\partial_{x_{N+1}}, \partial_\rho; \mathbf{n})$ is independent of the vector **n**, and $\hat{u}(x_{N+1}, \rho\mathbf{n})$ is a solution of the two-dimensional version of Eq. (5.14). If Eq. (5.14) is not invariant under $N$-dimensional rotations, then Eq. (5.16) is not its two-dimensional version. For example, consider, again, Eq. (2.12) for the electrostatic potential in an anisotropic medium. Writing $u(x,y,z)$ in terms of its Radon transform in $x$ and $y$ [with $\mathbf{n} = (n_x, n_y)$], one readily finds that $\tilde{P}$ is explicitly **n** – dependent:

$$\tilde{P}(\partial_{x_3}, \partial_\rho; \mathbf{n}) = \partial_z^2 + \left(\alpha n_x^2 + \beta n_y^2\right)\partial_\rho^2 \quad . \tag{5.19}$$

When $\alpha = \beta$, Eq. (2.12) is invariant under rotations in $R^2$, the operator $\tilde{P}$ does not depend on **n** explicitly, and is the two-dimensional version of the differential operator in Eq. (2.12).

When the original PDE is invariant under the rotation group in $\mathbf{x} \in R^N$, then Eq. (5.16) is the two-dimensional version of Eq. (5.14). The assumptions that $u(x_{N+1}, \mathbf{x})$ has an invertible Fourier integral representation in $\mathbf{x} \in R^N$, and that $u(x_{N+1}, \mathbf{x})$ and a sufficient number of its derivatives vanish at infinity, then lead to the result [6] that the solution in $(N+1)$ dimensions may be written as a superposition of solutions of the same equation in two dimensions:

$$u(x_{N+1}, \mathbf{x}) = \int_{|\mathbf{n}|=1} d\Omega_N \, F(x_{N+1}, \mathbf{n} \cdot \mathbf{x}; \mathbf{n}) \quad , \tag{5.20}$$

where **n** spans the $N$-dimensional unit sphere, and $\Omega_N$ is the angular variable over that sphere. $F(x_{N+1}, \rho; \mathbf{n})$ is obtained from $\hat{u}(x_{N+1}, \rho\mathbf{n})$ [and is, hence, also a solution of Eq. (5.16)] in a manner that depends on the parity of $N$:

$$F(x_{N+1}, \mathbf{n} \cdot \mathbf{x}; \mathbf{n}) = \begin{cases} \dfrac{1}{2(2\pi)^{N-1}} \left(\dfrac{-i\partial}{\partial \rho}\right)^{N-1} \hat{u}(x_{N+1}, \rho \mathbf{n}) \bigg|_{\rho = \mathbf{x} \cdot \mathbf{n}} & N \text{ odd} \\ i\mathcal{H} \dfrac{1}{2(2\pi)^{N-1}} \left(\dfrac{-i\partial}{\partial \rho}\right)^{N-1} \hat{u}(x_{N+1}, \rho \mathbf{n}) \bigg|_{\rho = \mathbf{x} \cdot \mathbf{n}} & N \text{ even} \end{cases}, \quad (5.21)$$

where

$$\mathcal{H}f(t) = \frac{1}{\pi} P \int_{-\infty}^{\infty} \frac{f(s)}{t-s} ds . \quad (5.22)$$

This result is demonstrated in [6] for the case of the wave equation, where $x_{N+1}$ is the time variable. It also applies to solutions of the Laplace equation, where $x_{N+1}$ is a spatial variable.

The conclusion of Section 2, that the lower-dimensional representation is not confined to two dimensions, can be also reached in the Radon transform approach if the analysis of [6] is applied to a *mixed Fourier-Radon* transform. Exploiting the splitting of $\mathbf{k} \in R^N$ of Eq. (2.5), define

$$u(x_{N+1}, \mathbf{x}) = \frac{1}{(2\pi)^{m/2}} \int_{-\infty}^{\infty} e^{i\overline{\mathbf{q}} \cdot \overline{\mathbf{x}}} \tilde{F}\big[(x_{N+1}, \hat{\mathbf{x}}; \overline{\mathbf{q}})\big] d^m \overline{\mathbf{q}} , \quad (5.23)$$

where

$$\tilde{F}(x_{N+1}, \hat{\mathbf{x}}; \overline{\mathbf{q}}) = \frac{1}{(2\pi)^{(N-m)/2}} \int_{-\infty}^{\infty} \tilde{u}(x_{N+1}, \hat{\mathbf{q}}, \overline{\mathbf{q}}) e^{i(\hat{\mathbf{q}} \cdot \hat{\mathbf{x}})} d^{N-m} \hat{\mathbf{q}} , \quad (5.24)$$

and $\tilde{u}(x_{N+1}, \hat{\mathbf{q}}, \overline{\mathbf{q}})$ is the coefficient in the Fourier integral of $u(x_{N+1}, \mathbf{x})$ in $\mathbf{x} \in R^N$. The mixed transform is defined as

$$\hat{F}(x_{N+1}, \rho \hat{\mathbf{n}}; \overline{\mathbf{q}}) = \int_{-\infty}^{\infty} \tilde{F}(x_{N+1}, \hat{\mathbf{x}}; \overline{\mathbf{q}}) \delta(\rho - \hat{\mathbf{x}} \cdot \hat{\mathbf{n}}) d^{N-m} \hat{\mathbf{x}} = \int_{(\xi \cdot \hat{\mathbf{n}})=0} \tilde{F}(x_{N+1}, \rho \hat{\mathbf{n}} + \xi; \overline{\mathbf{q}}) d^{N-m-1} \xi , \quad (5.25)$$

where $\hat{\mathbf{x}}, \hat{\mathbf{n}} \in R^{N-m}$, $\rho \in R$, $|\mathbf{n}| = 1$, $\xi \in R^{N-m-1}$. A representation of $u(x_{N+1}, \mathbf{x})$ as a superposition of solutions of the same PDE in $m+2$ dimensions is obtained.

## 6. Concluding comments

Several techniques have been developed in the literature for the derivation of the representation of solutions of linear PDE's, with constant coefficients, in dimensions higher than 2 as superpositions of solutions of the two-dimensional version of the same PDE. The unified approach using the Fourier integral representation of the solution has a few advantages. First is the simplicity that leads naturally to superpositions in terms of solutions in dimensions not confined to 2. Secondly, these lower dimensional solutions are either plane waves, or superpositions thereof, an observation that is not self-evident in the other methods. Thirdly, unlike in the other methods, it is not required that the Fourier integral representation is invertible. This is advantageous because functions for which Fourier inversion is possible in terms of tabulated functions are rare even for a single variable. They are rarer in two, let alone in higher dimensions. Finally, in the singularity distribution representations, Eqs. (1.7) and (1.9), there is a difference between even and odd dimensions in the singular behaviour of the solution at the origin. In the Radon transform approach, there is a difference between even and odd dimensions in the computation of $F(x_{N+1}, \mathbf{n} \cdot \mathbf{x}; \mathbf{n})$ [see Eqs. (5.21) and (5.22)]. In the Fourier integral representation, there is no apparent difference associated with the parity of the dimension.

Symmetry properties of the PDE determine the transformation group over the elements of which integration is performed in representing the solution as a superposition of solutions in the transformed, lower-dimensional space. The general solutions of the Laplace equation in the ($N$+1)-dimensional Euclidean space and the wave equation in (1+$N$) dimensions may be constructed as superpositions of their solutions in lower dimensions. Of particular simplicity is the representation in terms of the 2-dimensional, or of the (1+1)-dimensional equations, respectively, computed at the two-dimensional points [($z'$, $x'$) and ($t'$, $x'$), respectively]. These points are connected to the point where the multi-dimensional solution is evaluated by the transformations under which the differential operator is invariant [($N$+1)-dimensional rotations and (1+$N$)-dimensional Lorentz transformations, respectively].

Representing high-dimensional solutions of a PDE as a superposition of lower-dimensional solutions of the same equation may be useful in numerical algorithms: Finding the solution in a low dimension, one may compute the high-dimensional solution using a numerical integration technique, such as Monte Carlo.

**Acknowledgements**  Constructive comments by J. Schiff (Bar-Ilan University), and G. Burde and I. Rubinstein (Blaustein Institute) are acknowledged.

REFERENCES

1. Whittaker, E. T., Math. Ann. LVII, 333-355 (1902); Whittaker, E. T and G. N. Watson, *A Course of Modern Analysis*, Cambridge University press (1962).
2. Ockendon, J., S. Howison, A. Lacey and A. Movchan, *Applied Partial Differential equations*, Oxford University Press, Oxford (1999).
3. Courant, R and D. Hilbert, *Methods of mathematical Physics*, vol. II, Wiley (1965).
4. Radon, J., Ber. Verh. Sächs. Akad., **69**, 262-277 (1917).
5. John, F., *Plane waves and spherical means*, Wiley , New York (1955).
6. Ludwig, D., Comm. Pure. Appl. Math., 49-81 (1966).
7. Körner, T. W., *Exercises for Fourier analysis*, Cambridge University Press (1993).
8. Shaw, W. T., *Complex Mathematica*, Cambridge University Press (forthcoming, 2004).
9. Penrose R. and W. Rindler, *Spinors and space-time*, vol. 2, *Spinor and Twistor methods in space-time geometry*, Cambridge University Press (1984).
10. Gradshsteyn, I. S. and I. M. Rhyzhik, *Tables of Integrals, series and Products*, Academic Press, New York (9165).